\documentclass[preprint,showpacs,aps]{revtex4}
\usepackage{epsfig}

\newcommand{\be}{\begin{equation}}
\newcommand{\ee}{\end{equation}}
\newcommand{\ba}{\begin{eqnarray}}
\newcommand{\ea}{\end{eqnarray}}

\begin{document}

\title{Trajectory of test particle around a slowly rotating relativistic star
       emitting isotropic radiation}

\author{Jae-Sok Oh\footnote{e-mail: ojs@astro.snu.ac.kr}}

\affiliation{Department of Physics and Astronomy, FPRD, Seoul
National University, Seoul 151-742, Korea}

\author{Hongsu Kim\footnote{e-mail: hongsu@kasi.re.kr}}

\affiliation{Korea Astronomy and Space Science Institute, Daejeon
305-348, KOREA}

\author{Hyung Mok Lee\footnote{e-mail: hmlee@astro.snu.ac.kr}}

\affiliation{Department of Physics and Astronomy, FPRD, Seoul
National University, Seoul 151-742, Korea.}

\begin{abstract}
    We explored the motion of test particles near
slowly rotating relativistic star having a uniform luminosity.  In
order to derive the test particle's equations of motion, we made use
of the radiation stress-energy tensor first constructed by Miller
and Lamb \cite{ML96}. From the particle's trajectory obtained
through the numerical integration of the equations of motion, it is
found that for sufficiently high luminosity, ``suspension orbit"
exists, where the test particle hovers around at uniform angular
velocity in the same direction as the star's spin. Interestingly, it
turned out that the radial position of the ``suspension orbit" was
determined by the luminosity and the angular momentum of the star
alone and was independent of the initial positions and the specific
angular momentum of the particle. Also found is that there exist not
only the radiation drag but also ``radiation counter-drag'' which
depends on the stellar radius and the angular momentum and it is
this radiation counter-drag that makes the test particle in the
``suspension orbit" to hover around at uniform angular velocity
which is greater than that induced by the Lense-Thirring effect
(i.e., general relativistic dragging of inertial frame).
\end{abstract}



\maketitle


\newpage
\begin{center}
{\rm\bf I. INTRODUCTION}
\end{center}

   Astrophysical accretion flow onto massive or compact stars is one
of the major concerns in Astronomy and Astrophysics.  In the present
work, we explore the ground work for the full understanding of the
accretion flow onto highly luminous slowly rotating relativistic
stars.  In the current treatment of the accretion
process, the effect of radiation pressure on the inflow has not been
fully addressed. Therefore in the present work, we attempt to include
systematically the role of the radiation pressure in the accretion
process. To this end, we explore the effect of the radiation pressure
on the motion of a single test particle. Eventually we
hope that this elementary study of ours will be extended to the
case of actual accretion flow which can be thought of as consisting
of large number of constituent single particles.

   We now begin with the summary of the present status of the research in the
literature along this line. In their pioneering work, Abramowicz,
Ellis, and Lanza \cite{AEL90} demonstrated that radiation from the
highly-luminous non-rotating spherical massive star generates a
``critical point" above the stellar surface, but they limited the
motion of the particles to one-dimensional radial direction alone
and did not deal with the case of rotating central stars emitting
isotropic radiation. Later on, Miller and Lamb \cite{ML93} extended
one-dimensional motion to two-dimensional one and pointed out that
the trajectory of the particle is significantly affected by the
radiation from the non-rotating star if the luminosity of the star
is greater than $\sim 1\%$ of the Eddington luminosity. They,
however, confined themselves to the spherical symmetric spacetime
describing the non-rotation of the central star. Some time later,
they considered \cite{ML96} the effects of slow rotation of the
central star and constructed the radiation stress-energy tensor
describing the radiation field from the slowly rotating central
star. They, however, failed to notice the emergence of critical
radius (which corresponds to the critical point reported in
Abramowicz et al.\cite{AEL90}) as they considered the set-up in
which the luminosity is well below the Eddington's critical value.

   In the present work, in order to understand the effects of
the radiation pressure on the accretion onto the highly
luminous rotating relativistic stars in a rigorous and complete
manner, we employ the radiation stress-energy tensor which is
given by Miller and Lamb \cite{ML96} as an elaboration on that given originally by Abramowicz et
al. \cite{AEL90} to describe the radiation emitted from the slowly
rotating central stars, and derive the equation of motion.

   By integrating numerically the
equations of motion, we realized that there exists the ``suspension
orbit" (where the test particle hovers around the central star) that
corresponds to the ``critical point" in \cite{AEL90}. And it turns
out that the radial position of this ``suspension orbit" depends on
the luminosity and the angular momentum of the central star alone,
and does not depend on the initial position and the initial angular
momentum of the test particles.

   In addition, it is realized that the test particle in the
``suspension orbit" has uniform azimuthal velocity in the same
direction as the star's spin motion.  It is interesting to note that
the uniform azimuthal velocity of the test particle at the
``suspension orbit" is greater than that due to the Lense-Thirring
effect.

   In section 2, we provide the derivation of the equations of
motion from the known radiation stress-energy tensor and discuss the
usual radiation drag terms and the newly discovered radiation
counter-drag terms.  In section 3, we present the results of the
numerical integrations showing the emergence of the ``suspension
orbit". In section 4, we analyze the motion of the particle in the
``suspension orbit", and finally, in section 5, we end with the
discussion of our results.

\begin{center}
{\rm\bf II. EQUATIONS OF MOTION}
\end{center}

To derive the equations of motion that govern the trajectory of the
test particles in the presence of radiation from the slowly rotating
relativistic star, we assume that radiation source emits
isotropically the radiation from the whole stellar surface and we
shall employ the following metric obtained from the Kerr black hole
metric (\cite{K63}) by retaining its terms to the first order in
Kerr parameter $a$ to represent the spacetime exterior to the slowly
rotating star,
\begin{eqnarray}
   ds^2 &=& g_{\mu\nu}dx^{\mu}dx^{\nu} \nonumber \\
        &=& -(1- 2M/r)dt^{2} \,+\, (1- 2M/r)^{-1}dr^{2} \,+\, r^2 d\theta^{2}
        \,+\, r^2 {\sin}^{2}\theta (d\phi^{2} - 2\omega d\phi dt)
\end{eqnarray}
where $\omega = 2J/r^3$ is the Lense-Thirring angular velocity
(\cite{LT18}) arising due to the frame-dragging effect of a
stationary axisymmetric spacetime, whch can be identified with the
orbital angular velocity of the LNRF(Locally Non-Rotating Frame; see
\cite{B70}; \cite{BPT72}), $M$ and $J$ are the gravitational mass
and angular momentum of the star respectively. We work in the
geometric units where $G = c = 1$ (G is the gravitational constant
and c is the speed of light). Following Miller and Lamb \cite{ML96},
we introduce a dimensionless angular momentum $j \equiv cJ/(GM^2)$
and dimensionless velocity $v \equiv <v^{\hat\phi}>/c $ as a
convenient measure of the rotation rate of the star and the rotation
rate of the radiation source, respectively, where $<v^{\hat\phi}>$
is the appropriate average (see \cite{ML96}) of $v^{\hat\phi}$
(azimuthal linear velocity) over the emitting surface visible from
the test particle. The hat denotes the physical quantity measured in
the LNRF.  In this paper `slow rotation' means $j \ll 1$ and $v \ll
1$ and we keep terms that are only first-order in $j$ and $v$.  For
a neutron star with radius $R \approx 10$ km, mass $M \approx 1.4
M_{\bigodot}$, and spin frequency $\nu_{s} \approx 600$ Hz(i.e., a
millisecond pulsar), the dimensionless angular momentum $j$ is
approximately 0.2.

    The equations of motion (which actually is the geodesic equation) are given by
\begin{equation}
   a^{\alpha} = \frac {f^{\alpha}}{m}
\end{equation}
where $f^{\alpha}$ denotes a radiation force exerted by the
radiation (or luminosity) on the test particle, $m$ is the rest mass
of the particle, and
\begin{equation}
  a^{\alpha} = \frac {dU^{\alpha}}{d\tau} \; + \;
  \Gamma^{\alpha}_{\mu\nu} U^{\mu}U^{\nu}
\end{equation}
is the acceleration, with $U^{\alpha}$ being the four-velocity of the
particle, $\Gamma^{\alpha}_{\mu\nu} = \frac{1}{2} g^{\alpha\beta}
(g_{\beta\mu ,\nu} + g_{\beta\nu ,\mu} - g_{\mu\nu ,\beta})$ being the
Affine connection, and comma (,) denoting partial derivatives.

    For the sake of computational convenience, like in \cite{ML96},
we assumed that the radiation scatters off the test particles and
the momentum-transfer cross section $\sigma$ of the test particle is
independent of energy (frequency) and direction of the radiation.
Hence the radiation force $f^{\alpha}$ due to scattering of the radiation
 is proportional to and in the direction of the
radiation flux in the comoving frame (particle's rest frame), and is
given by (see \cite{LM95})
\begin{equation}
   f^{\alpha} = \sigma F^{\alpha},
\end{equation}
where $F^{\alpha}$ is the quantity obtained by transforming the
radiation energy flux $T_{co}^{\hat i\hat 0}$ measured in the
comoving frame using the orthonormal tetrad $\tilde
{e}^{\alpha}_{\hat i}$ associated with the particle's rest frame as
follows,
\begin{eqnarray}
   F^{\alpha} &=& \tilde {e}^{\alpha}_{\hat i} T_{co}^{\hat i\hat0}
               \nonumber \\
              &=& - h^{\alpha}_{\beta} T^{\beta\sigma} U_{\sigma},
\end{eqnarray}
where $h^{\alpha}_{\beta} = \delta^{\alpha}_{\beta} +
U^{\alpha}U_{\beta}$ is the projection tensor that projects onto
each spacelike hypersurface and $T^{\beta\sigma}$ is the radiation
stress-energy tensor in first-order (in $j$) Boyer-Lindquist
coordinates(see \cite{ML96})(for $j \neq 0$).

    According to \cite{ML96}, the components of the radiation stress-energy tensor
$T^{\hat\alpha \hat\beta}$ as measured in LNRF are given by,
\begin{eqnarray}
   T^{\hat t\hat t} &\approx& 2\pi I_{0}(r)(1-\cos\alpha_{0})
        \\ \nonumber
   T^{\hat t\hat r} &\approx& \pi I_{0}(r)\sin^{2}\alpha_{0}
        \\ \nonumber
   T^{\hat t\hat\phi} &\approx& \frac{\pi}{3} I_{0}(r)\mathcal{J}(r)
        (\cos^{3}\alpha_{0} - 3\cos\alpha_{0} + 2)
        \\ \nonumber
   T^{\hat r\hat r} &\approx& \frac{2\pi}{3} I_{0}(r)(1-\cos^{3}\alpha_{0})
        \\ \nonumber
   T^{\hat r\hat\phi} &\approx& \frac{\pi}{4}
        I_{0}(r)\mathcal{J}(r)\sin^{4}\alpha_{0}
        \\ \nonumber
   T^{\hat\theta \hat\theta} &\approx& \frac{\pi}{3} I_{0}(r)
        (\cos^{3}\alpha_{0} - 3\cos\alpha_{0} + 2)
        \\ \nonumber
   T^{\hat\phi \hat\phi} &\approx& \frac{\pi}{3} I_{0}(r)
   (\cos^{3}\alpha_{0} - 3\cos\alpha_{0} + 2) \nonumber,
\end{eqnarray}
where the subscript 0 denotes the quantity for the case of
non-rotating star, $\alpha_{0}$ is an apparent viewing angle of the
star seen by a locally static observer in Schwarzschild spacetime
and is given by $\sin\alpha_{0} = \left(\frac{R}{r} \right) {\left(
\frac{1 - 2M/r}{1 - 2M/R}\right)}^{1/2}$ (see \cite{AEL90}) for the
radius of the star $R \geq 3M$, and $I_{0}(r)$ is the
frequency-integrated specific intensity at the radial position $r$
and is given by (see Appendix A in \cite{ML96})
\begin{eqnarray}
   I_{0}(r) = \frac{(1-2M/R)}{(1-2M/r)^2} \frac{m M}{\pi\sigma R^2}
          \left(\frac{L^{\infty}}{L_{Edd}^{\infty}}\right),
\end{eqnarray}
where the Eddington luminosity $L^{\infty}_{Edd} \equiv 4\pi m
M/\sigma$ is the luminosity of a spherically symmetric source such
that at infinity the outward radiation force  balances the inward
gravity (see \cite{LM95}) and $L^{\infty}$ is the
luminosity of the star as measured by an observer at infinity, and
$\mathcal{J}(r)$ is given by
\begin{eqnarray}
   \mathcal{J}(r) = 8j\left(\frac{r}{M}\right)\left(\frac{M^3}{R^3}-\frac{M^3}{r^3}\right)
          + 4v\left(\frac{r}{R}\right)(1-2M/R)^{1/2}.
\end{eqnarray}

    By transforming the above radiation stress-energy tensor $T^{\hat\mu \hat\nu}$
to the LNRF using tetrad $e^{\alpha}_{\hat\mu}$ (which is given
below) associated with the LNRF, the radiation stress-energy tensor
$T^{\alpha\beta}$ in the first-order Boyer-Lindquist coordinates
(see \cite{ML96}) is obtained as,
\begin{eqnarray}
   T^{\alpha\beta} = e^{\alpha}_{\hat\mu} e^{\beta}_{\hat\nu} T^{\hat\mu\hat\nu},
\end{eqnarray}
where the tetrad associated with the LNRF are,
\begin{eqnarray}
   e^{\hat 0} &=& (1 - 2M/r)^{1/2}dt, \nonumber \\
   e^{\hat 1} &=& (1 - 2M/r)^{-1/2}dr, \\
   e^{\hat 2} &=& rd\theta, \nonumber \\
   e^{\hat 3} &=& -2j\frac{M^2}{r^2}\sin\theta + r\sin\theta d\phi. \nonumber
\end{eqnarray}

    We now focus our attention on
the orbits confined to the equatorial plane ($\theta =
\frac{\pi}{2}, U_{\theta} = 0$). The decomposition into each
component of the equations of motion (5) in tensor form is given in
the Appendix.  It should also be noted that as the radiation
stress-energy tensor (equation (6)) first constructed by Miller and
Lamb \cite{ML96} is valid up to $j=0.20$, the equations of motion
derived from it also valid within above ranges.

  Since background spacetime of equation (1) has a rotational
isometry, Killing theorem states that there exists a rotational
Killing field $\eta^{\mu} = \delta^{\mu}_{\phi}$ such that the test
particle's specific angular momentum $l =
g_{\mu\nu}\eta^{\mu}U^{\nu} = g_{\mu\nu}\delta^{\mu}_{\phi}U^{\nu} =
g_{\phi\nu}U^{\nu} = U_{\phi}$ is conserved. Therefore, the
azimuthal component (equation (23)) of the equations of motion in
the Appendix governs the time evolution of the test particle's
specific angular momentum and can be rewritten as,

\begin{eqnarray}
   \frac{dU_{\phi}}{d\tau} &=& -\frac{L}{3}\frac{f(r)}{(1-2M/r)}
                                \left[ A(\alpha_{0}) U_{t}^2
                              + \left(1-\frac{2M}{r}\right)^2
                               (\cos\alpha_{0}\sin^2\alpha_{0})U_{r}^2 \right] U_{\phi}
                                \nonumber \\
                           &-&  \frac{L}{3}\frac{f(r)}{(1-2M/r)}
                                \left[ 4j\left( \frac{M^2}{r^3} \right)
                                A(\alpha_{0}) U_{\phi}^2 + r \mathcal{J}(r)B(\alpha_{0})
                              + \left( \frac{2}{r} \right)\mathcal{J}(r)B(\alpha_{0}) U_{\phi}^2
                                \right] U_{t} \nonumber \\
                           &-&  L f(r) \left[ (2\sin^2\alpha_{0})U_{t}U_{\phi}
                              + 4j\left( \frac{M^2}{r^3} \right)
                              (\sin^2\alpha_{0} )U_{\phi}^2 \right] U_{r}
                                \nonumber \\
                           &-&  L f(r)
                                \left[ \left(\frac{r}{4} + \frac{1}{2r} U_{\phi}^2 \right)
                                \right] \mathcal{J}(r) (\sin^4\alpha_{0}) U_{r},
\end{eqnarray}
where $A(\alpha_{0}) = \cos^3\alpha_{0} -9\cos\alpha_{0} + 8$,
$B(\alpha_{0}) = \cos^3\alpha_{0} -3\cos\alpha_{0} + 2$,  $L \equiv
\left(\frac{L^{\infty}}{L_{Edd}^{\infty}}\right)$ is the luminosity parameter,
and $ f(r) = \frac{M}{R^2} \frac{(1-2M/R)}{(1-2M/r)^2}$.

We are now ready to envisage the features of the solution to the
azimuthal component of the equation of motion given in equation
(11). Obviously, however,  the analytic solution to this coupled
nonlinear ordinary differential equation (11) is not readily
available. Therefore, we shall look for its numerical solution in
the next section. However, even before that we can read off the
essential features of the solution. To this end, we will interpret
this azimuthal component of the equation of motion as the equation
that determines the time evolution of the test particle's specific
angular momentum as $U_{\phi} = l$.  To summarize, this azimuthal
component of equation of motion breaks into three parts: the first
part is the line one of the equation (11) and the two terms in this
line are responsible for the radiation drag, that is, the well-known
Poynting Robertson effect since it is manifest that the overall sign
of this line is negative definite and these two terms are linearly
proportional to the test particle's specific angular momentum
$U_{\phi}$ and the star's luminosity. The second part is the line
two of the equation (11) and three terms in this line are
responsible for the radiation counter-drag, that has the effect
opposite to the Poynting Robertson effect since it is obvious that
the overall sign of the line two is positive definite. Third part
consists of line three and line four of the equation (11) and terms
with opposite signs can be regarded as being responsible for, say,
radiation drag for one sign and radiation counter-drag for the
other. Now, in what follows, let us be more specific on the nature
of each term in these lines.  For instance, the first term in line
two is due to the Lense-Thirring effect, that is, the dragging of
inertial frame and the second and third terms in line two are due to
the spin of the central star. Next, second term in line three is due
to the Lense-Thirring effect and two terms in line four are due to
the spin of the central star.

It is interesting to note that although the radiation drag or the
Poynting Robertson effect has been long-known, the nature of this
effect has not been unveiled manifestly thus far. In equation (11)
above, however, we were successful to quantify this Poynting
Robertson effect by explicitly identifying the terms in line one
which are responsible for the effect. In other words, by examining
the equation of motion of a test particle placed in the background
of luminous relativistic spinning star, we demonstrated manifestly
that actually the Poynting Robertson effect takes place.
Interestingly, however, this is not the end of the story.
Remarkably, in the test particle's equation of motion, there are
also terms which appear to be responsible for the effect just
opposite to the Poynting Robertson effect, that is, terms in line
two and some more terms in line three and four.  This last point
appears to imply that particularly for  luminous relativistic
spinning star, the new effect just opposite to the Poynting
Robertson  effect, which will be coined henceforth as the
``radiation counter-drag", takes place as well.  To the best of our
knowledge, the counter-drag effect of this sort has never been
reported in the literature so far. Therefore, in the following
sections, we will solve the test particle's equation of motion
numerically to construct and investigate quantitative solutions that
will support our analysis of the features of the solutions stated
above.  What is more, based on both this numerical analysis and
analytical approach, we will report on the emergence of the
``suspension orbit" (which turns out to be the extension of the
critical point pointed out in earlier study \cite{AEL90} in the
absence of star's rotation).

\begin{center}
{\rm\bf III. NUMERICAL INTEGRATION}
\end{center}
In this subsection, we shall present the numerical solution to the equations of motion derived in section 2.
We begin with the brief description of our treatment of this numerical analysis.
Using the  equations of motion derived in section 2, we have followed
the trajectory of the particle in the presence of radiation from the
slowly rotating star. We confine the motion of the particle onto the
equatorial plane so that the polar angle component of the velocity
$U_{\theta}$ is set to be zero ($U_{\theta} = 0$). We assume that
the star has uniform density, so the angular velocity of the star
and the angular velocity of the LNRF at the stellar surface are
given, respectively, by $\Omega = \frac{5}{2}
j\left(\frac{M}{R^2}\right)$ and $\omega =
2j\left(\frac{M^2}{R^3}\right)$.  Thus, the average azimuthal
velocity $v^{\hat\phi}$ of the radiation source as measured by an
observer in the LNRF is calculated to be,
\begin{eqnarray}
      v = \frac{1}{\pi} j \left(\frac{M^2}{R^2}\right)
          \left(1-\frac{2M}{R}\right)^{-1/2}
          \left[5\left(\frac{R}{M}\right)-4\right]. \nonumber
\end{eqnarray}

\begin{figure}
\centerline{\epsfig{file=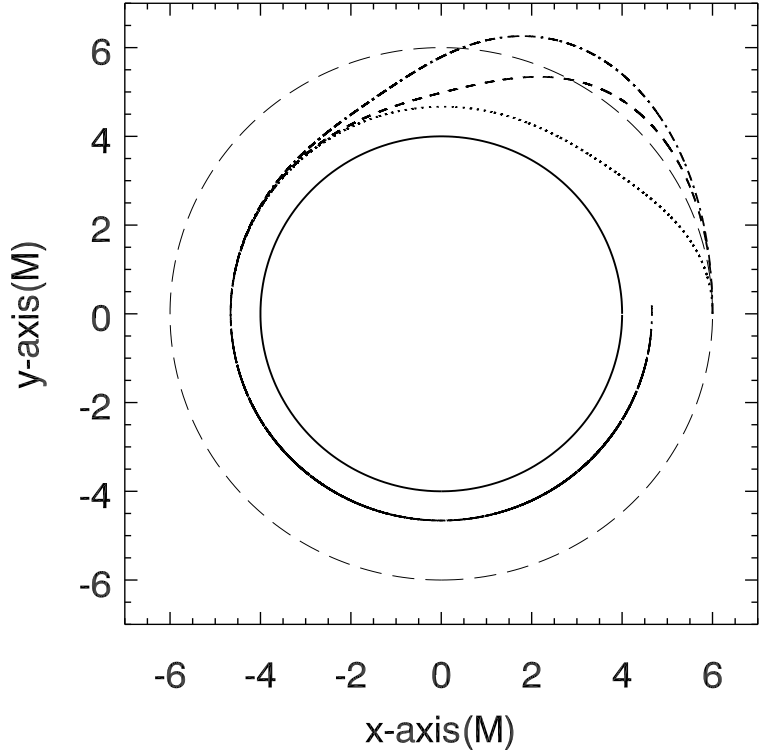, width=10cm, height=10cm}}
\caption{shows the trajectories of the particles having three
azimuthal velocities of 0.10(dotted curve), 0.25(dashed curve), and
0.30(dash-dotted curve), respectively when they initially co-rotate
with the star(which is rotating counter-clockwise all the way)
having angular momentum $j=0.1$.}
\end{figure}

\begin{figure}
\centerline{\epsfig{file=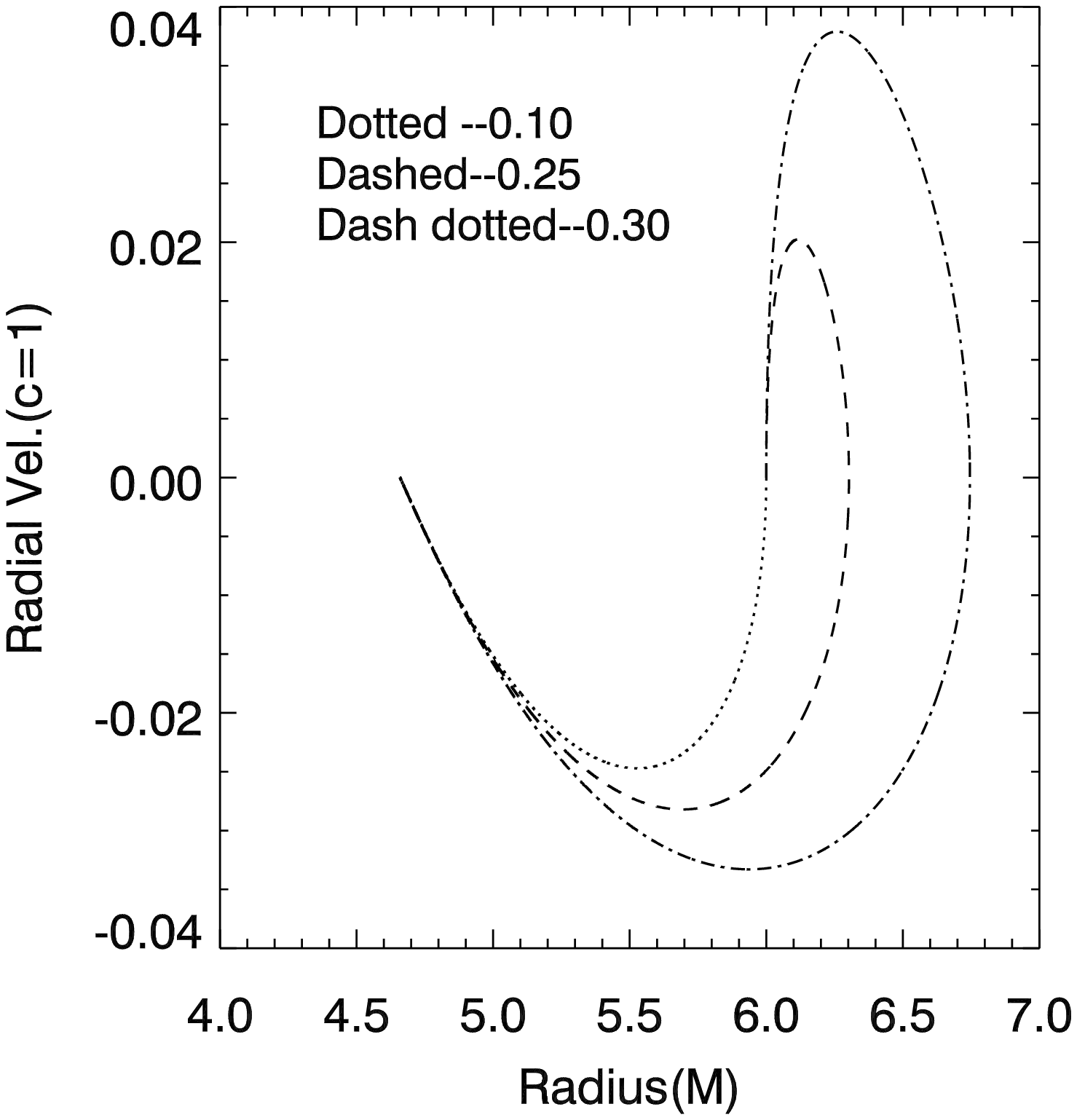, width=10cm, height=10cm}}
\caption{shows locally measured radial velocities as functions of
radius for particles in Fig. 1 inflowing from $r=6M$ toward the
``suspension orbit".}
\end{figure}

\begin{figure}
\centerline{\epsfig{file=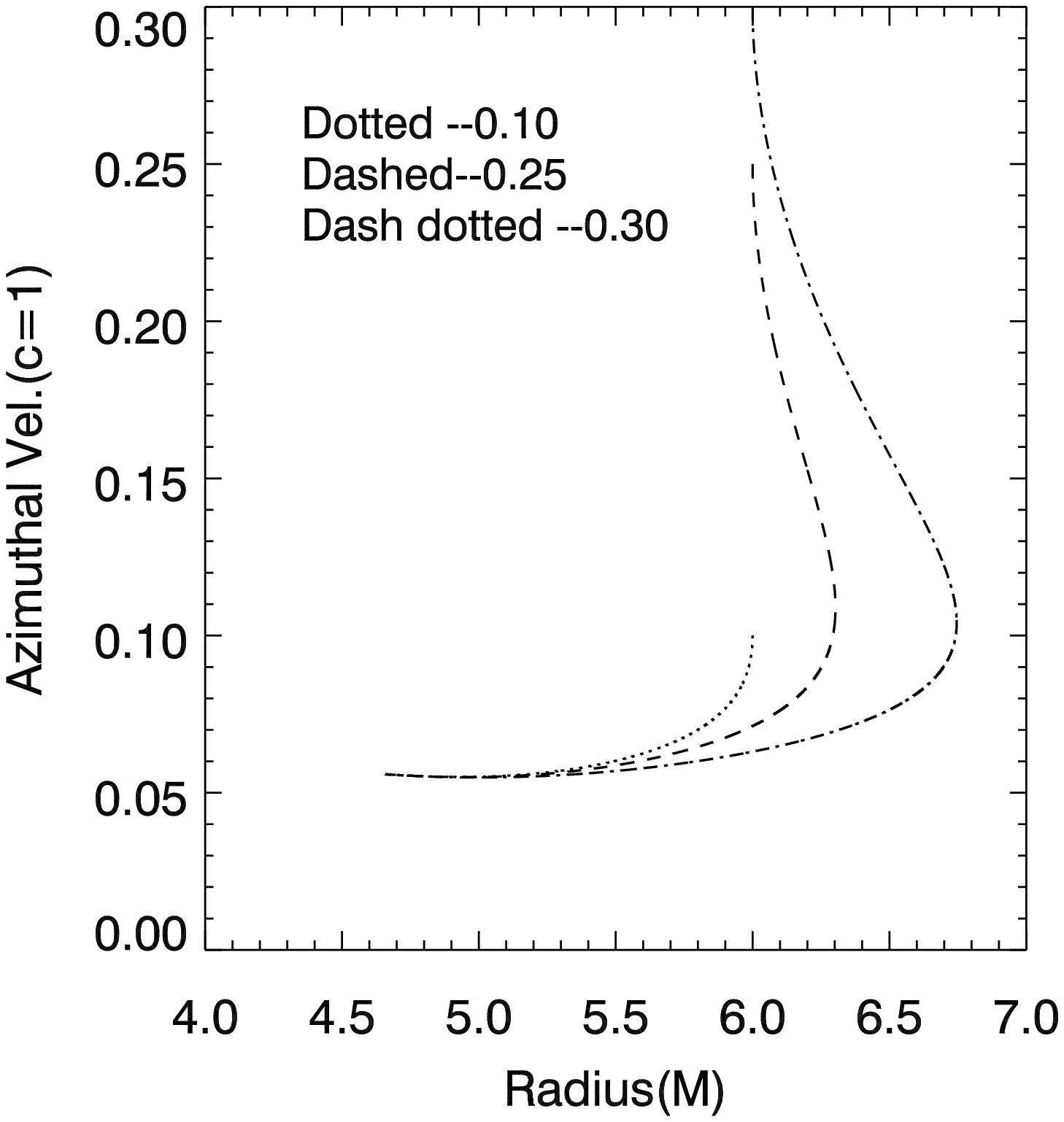, width=10cm, height=10cm}}
\caption{shows locally measured azimuthal velocities as functions of
radius for particles in Fig. 1 inflowing from $r=6M$ toward the
``suspension orbit".}
\end{figure}

   Fig. 1 shows the trajectories of the particles starting at the
position of $r = 6M$ where the particles have three azimuthal
velocities of 0.10(dotted curve), 0.25(dashed curve), and
0.30(dash-dotted curve), respectively. Initial radial velocities of
all the particles are equal to zero and the luminosity parameter of
the star with $R=4M$ is $L=0.75$. Solid line with
radius of $4M$ denotes the stellar surface, long dashed circle is
the virtual circular orbit with radius of $r=6M$ (i.e., ISCO;
Innermost Stable Circular Orbit). The starting point of all the
particles are the same as $(6M, 0)$ in Cartesian coordinates $(x,
y)$. The rotation of the star is counter-clockwise. Hence the
particles start counter-clockwise rotational motion.

   As can be seen in Fig. 1, although the three particles start out
in different azimuthal velocities, they end up being along the same
circular orbit which we henceforth shall refer to the ``suspension
orbit". It turns out that this ``suspension orbit" lies in between
the stellar surface(i.e., $r=4M$) and the ISCO(i.e., $r=6M$).
According to the numerical analysis, its radius is given by $r =
4.66M$.

   Fig. 2 shows the radial velocities of the particles in Fig. 1
inflowing toward the stellar surface from $r=6M$ as a function of
the radius. Numerical integration demonstrates that the radial
velocities of the particles at the ``suspension orbit" vanish and
the time rate of change of the radial velocities is also equal to
zero, i.e., $U_{r}=0$ and $\frac{dU_{r}}{d\tau} = 0$ at the
``suspension orbit".

   Fig. 3 shows the azimuthal velocities of the particles in Fig. 1 as
a function of the radius, and there we can figure out that the
azimuthal velocities $v^{\tilde\phi}$ of the particles at the
``suspension orbit" as measured by a locally static observer are all
equal to $v^{\tilde\phi} \simeq 0.056 $ in units of c(the speed of
light) and the time rate of change of the azimuthal velocities is
equal to zero, i.e., $U_{\phi}=\rm constant$ and
$\frac{dU_{\phi}}{d\tau} = 0$ at the ``suspension orbit". These
indicate that the particles orbit at constant speed there.
   Remarkably, essentially the same is true for the case where the
central star is not rotating.  This point appears to indicate that
the emergence of both the critical point for the case of
non-rotating central star and the ``suspension orbit" for the
present case of rotating central star are indeed a generic feature
that a high luminosity relativistic star exhibits.
\\

   Thus far in Figs. 1 through 3, we have studied the case where the
central star and the test particles ``co-rotate" all the way.  Next,
we move on to the other case where they counter-rotate initially but
end up co-rotating eventually and here, the rotation of the star is counter-clockwise all the way. 
The result is given in Figs. 4
through 6.

   To summarize, in Figs. 1 through 3, we have studied the case where the
central star and the test particles ``co-rotate" all the way. In
Figs. 4 through 6, however, we have studied the other case where
they counter-rotate initially but end up co-rotating eventually.
Namely, regardless of initial conditions, i.e., whether they are
initially co-rotating or counter-rotating, the system reaches the
same final equilibrium state where the test particles end up
co-rotating with the central star. This result indeed is very
interesting and curious particularly for the case when the central
luminous star is spinning since the eventual fate of the test
particles is the co-rotation with the central star at the
``suspension orbit". Therefore, we need careful understanding of the
underlying physics and our interpretation is based upon the geodesic
equation of the test particles at the ``suspension orbit" given in
equation (13) below and it can be describes as follows.

Firstly, first line term in equation (13) plays the role of radiation drag
because its overall sign is negative definite due to $U_{t} < 0$ and
is proportional to the test particle's specific angular momentum
$U_{\phi}$. Note that the radiation drag of the first line term
works on the test particle regardless of the spin of the central star.

Secondly, fourth line terms, on the other hand, play the role of ``radiation counter-drag''
because its overall sign is positive definite and
is proportional to the central star's angular momentum
$j$. Note that the radiation counter-drag of these fourth line terms
work on the test particle regardless of the test particle's specific angular momentum
$U_{\phi}$.

Indeed, the radiation drag term in the first line of equation (13)
that represent the Poynting-Robertson effect is well-known and it is
generic as it is independent of the central star's angular momentum.
The radiation counter-drag terms in the fourth line has been
neglected in our conventional understanding of the
Poynting-Robertson effect and hence is rather unfamiliar.  Besides,
it is not so generic as it appears only in the presence of the
central star's angular momentum.  Therefore, unlike the case where
the central luminous star is non-rotating, for the case at hand
where the central star is spinning, not only the radiation drag term
in the first line but the radiation counter-drag terms in the fourth
line operate as well and indeed they would affect the motion of the
test particle on equal footing to determine its trajectory.
Interestingly enough, our numerical study above exhibits that once
the test particle arrives at the ``suspension orbit", it never comes
to a complete stop.  Rather, it keeps co-rotating with the central
luminous spinning star and this is the new discovery of the present
work which has not been realized in the previous literature
addressing the similar related issues.

\begin{figure}
\centerline{\epsfig{file=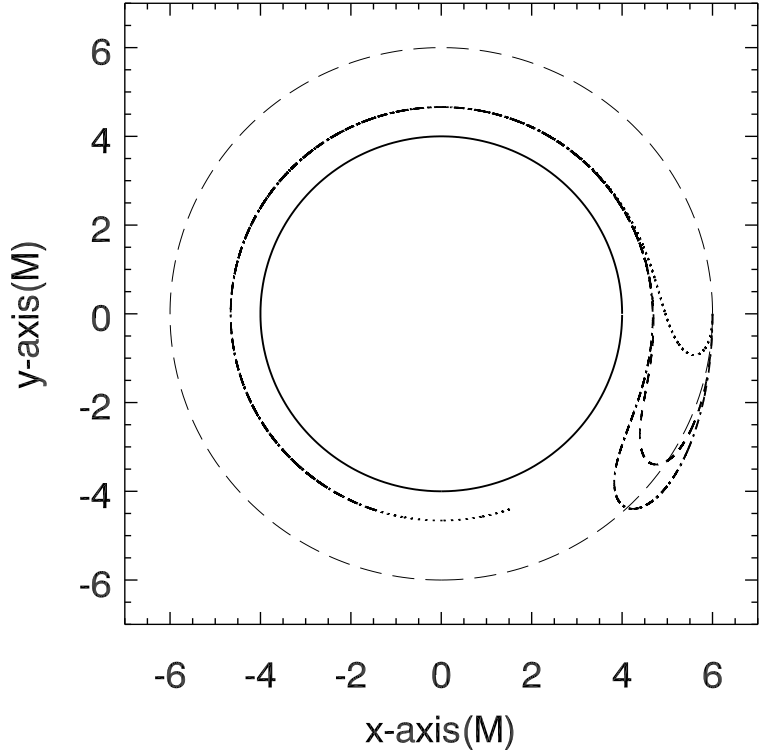, width=10cm, height=10cm}}
\caption{shows the trajectories of the particles having three
azimuthal velocities of -0.10(dotted curve), -0.25(dashed curve),
and -0.30(dash-dotted curve), respectively when they initially
counter-rotate with the star(which is rotating counter-clockwise all
the way) having angular momentum $j=0.1$ but end up being
co-rotating with the star mainly due to the radiation counter-drag.}
\end{figure}

\begin{figure}
\centerline{\epsfig{file=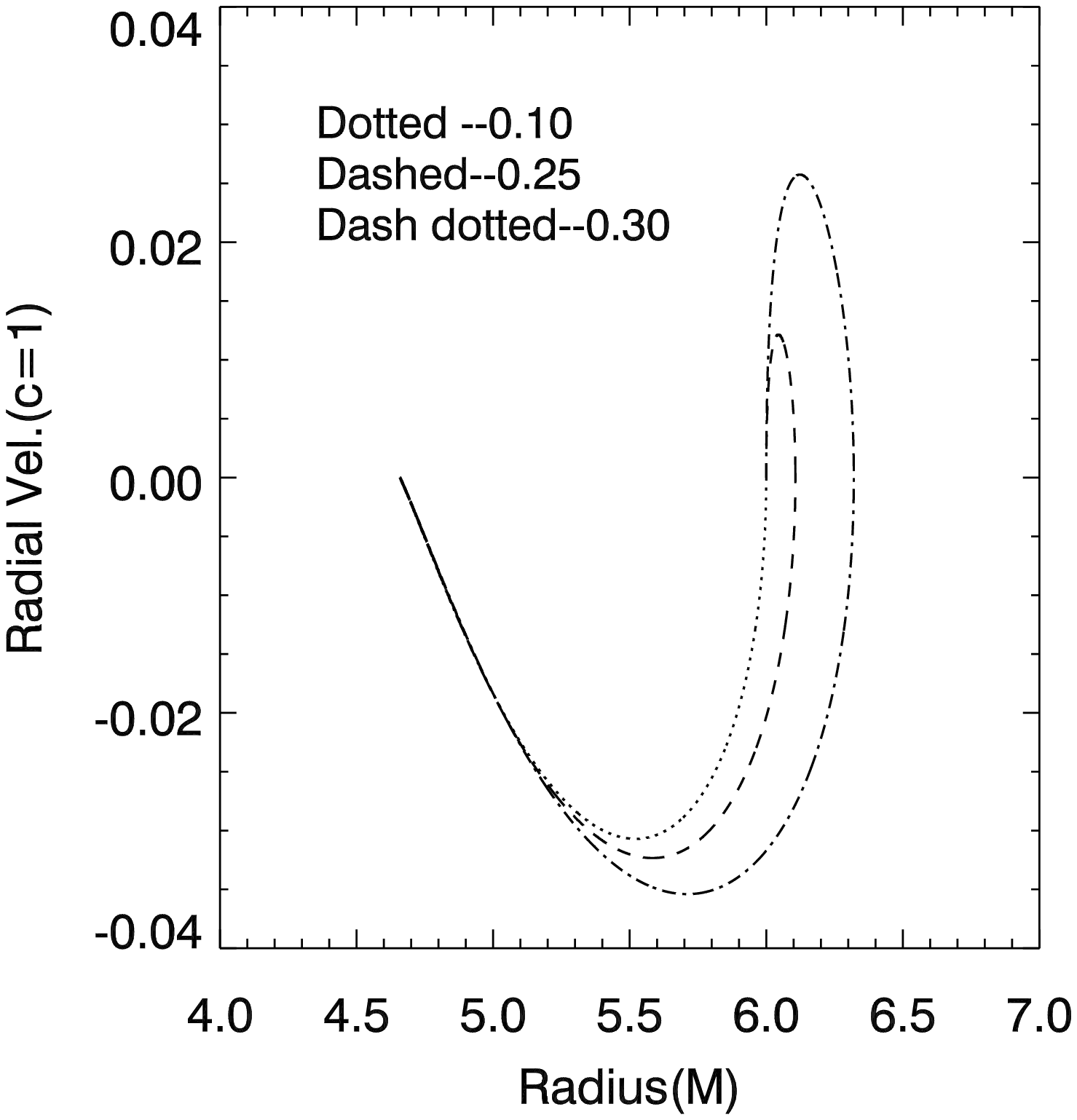, width=10cm, height=10cm}}
\caption{shows locally measured radial velocities as functions of
radius for particles in Fig. 4 inflowing from $r=6M$ toward the
``suspension orbit".}
\end{figure}

\begin{figure}
\centerline{\epsfig{file=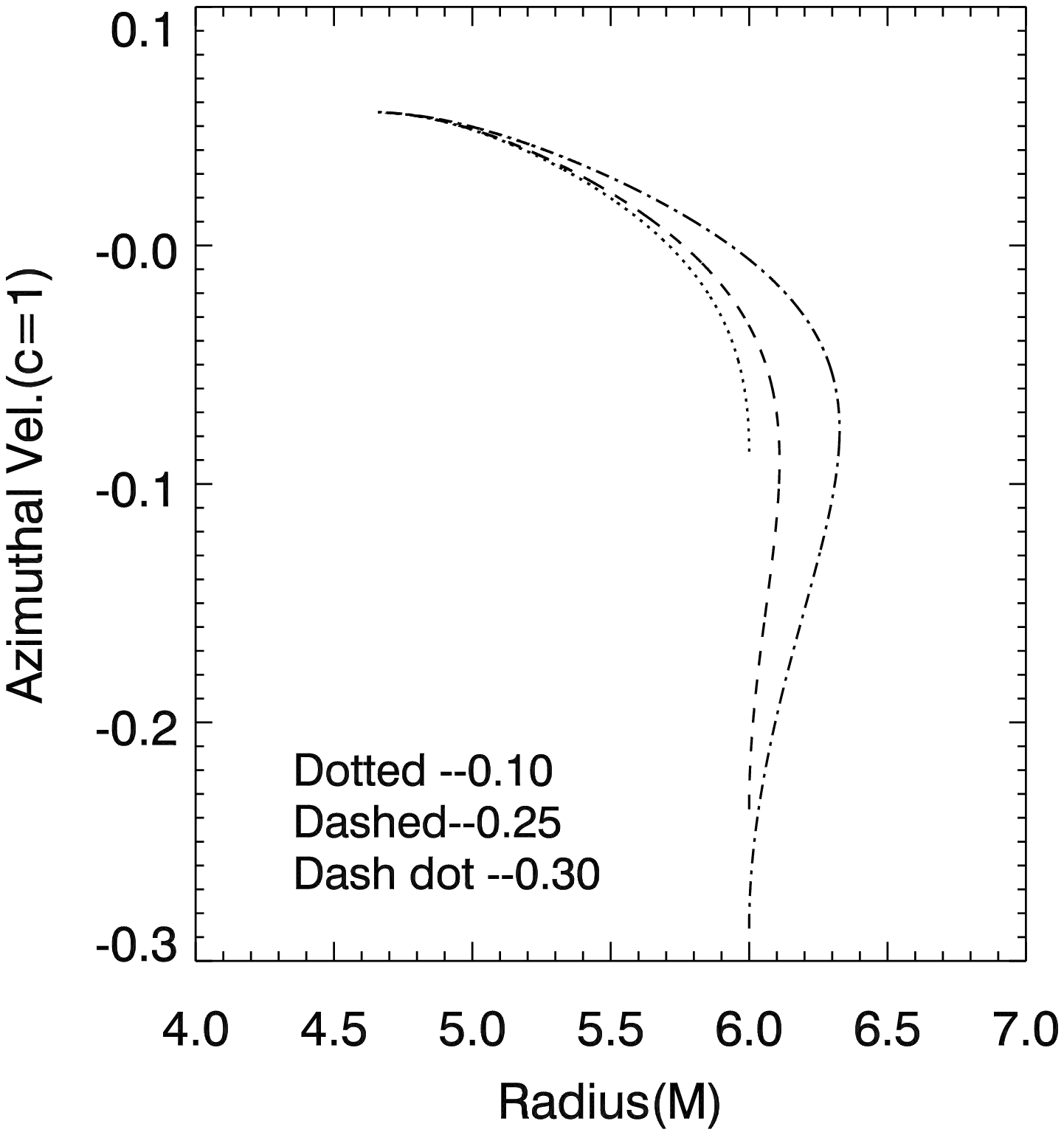, width=10cm, height=10cm}}
\caption{shows locally measured azimuthal velocities as functions of
radius for particles in Fig. 4 inflowing from $r=6M$ toward the
``suspension orbit".}
\end{figure}

   As can be noticed from the numerical
integration, the radiation from the slowly rotating star makes the
test particles to hover around the star with uniform azimuthal
velocity regardless of the initial position and the initial  angular
momentum of the particle, and the particle's motion in the
``suspension orbit" is characterized by the following conditions,
\begin{eqnarray}
   U_{r} &=& 0 \nonumber \\
   U_{\phi} &=& \rm {constant} \nonumber \\
   \frac{dU_{r}}{d\tau} &=& \frac{dU_{\phi}}{d\tau} =0. \nonumber
\end{eqnarray}

\begin{center}
{\rm\bf IV. EXAMINATION OF THE MOTION IN THE ``SUSPENSION ORBIT"}
\end{center}

   In order to understand the nature of ``forces" exerted on the particle
hovering around the ``suspension orbit", let us examine the
equations of motion using the conditions mentioned above in section
3 and then determine the coordinate radius of the ``suspension
orbit" and the azimuthal velocity of the particle as measured by the
locally static observer.

    By inserting $U_{r} = 0$ and
$\frac{dU_{\phi}}{d\tau} =0$ of the conditions for the ``suspension
orbit" into the $\phi$-component (equation (11)) of the equations of
motion, we obtain the following,
\begin{eqnarray}
       \left[4j\frac{M^2}{r^2}A(\alpha_{0}) + 2\mathcal{J}(r)B(\alpha_{0})\right]
       U_{\phi}^2
       + r A(\alpha_{0})U_{\phi}U_{t}
       + r^2 \mathcal{J}(r) B(\alpha_{0}) = 0,
\end{eqnarray}
where the luminosity parameter $L$ as a common factor is omitted.
Then, plugging $\mathcal{J}(r)$ in equation (8) into the above
equation (12) gives, after some manipulation,
\begin{eqnarray}
       0 = \frac{dU_{\phi}}{d\tau} &=& A(\alpha_{0})U_{t}U_{\phi}
               \nonumber \\
         &+& 4j\left(\frac{M^2}{r^3}\right)
               \left[A(\alpha_{0})-4B(\alpha_{0})\right]U_{\phi}^2
               \nonumber \\
         &+& 8B(\alpha_{0})\left[2j\left(\frac{M^2}{R^3}\right)
              + v\left(\frac{1}{R}\right)(1-2M/R)^{1/2}\right]U_{\phi}^2
               \nonumber  \\
         &+& 8j\left( \frac{r^2}{R^3} - \frac{1}{r} \right) M^2 B(\alpha_{0})
         + 4v\left( \frac{r^2}{R} \right) \left( 1- 2M/R \right)^{1/2} B(\alpha_{0})
\end{eqnarray}

   Firstly, terms in line one in equation (13) play the role of radiation drag
because its overall sign is negative definite due to $U_{t} < 0$ and
is proportional to the test particle's specific angular momentum
$U_{\phi}$. Therefore, the radiation drag terms in line one
work on the test particle having the azimuthal velocity $U_{\phi}$
regardless of the spin of the central star.
   Secondly, as $\left[A(\alpha_{0})-4B(\alpha_{0})\right]$
of line two in equation (13) has positive value for $\alpha_{0} \neq
0$, the overall sign of line two is positive definite, thus this
line serves as the radiation counter-drag that speeds up the
azimuthal motion.  It should also be noted that this line two is
proportional to the Lense-Thirring angular velocity $\omega =
2j\left(\frac{M^2}{r^2}\right)$ and the test particle's specific
angular momentum square $U^{2}_{\phi}$, thus the radiation
counter-drag of this libe two is due to the Lense-Thirring effect
arising from the rotation of the central star. Unfortunately, this
line two in equation (13) has very tiny contribution to the
azimuthal velocity of the particle because of $U_{\phi} \sim j \ll
1$ in the ``suspension orbit". This proportionality relation between
$U_{\phi} $ and $j$ can be noticed from the fact that the
``suspension orbit" in slow rotation case amounts to critical point
(see \cite{AEL90}) in non-rotating case and $U_{\phi}$ at the
critical point is equal to zero, and also can be supported by
subsequent calculation in equation (15). Therefore, all terms
including $U_{\phi}^2$ is negligible in the ``suspension orbit".
However, in the case where it is not allowed to ignore $U_{\phi}$
like in initial trajectory at $r=6M$, all the radiation counter-drag
terms including $U_{\phi}^2$ could contribute somewhat to the
azimuthal velocity of the particle.
   Thirdly, two terms in line three in equation (13) have overall positive definite
sign, and thus line three behaves as the radiation counter-drag.
Since this line three is proportional to the cental star's angular
momentum $j$, the radiation counter-drag of line three is due to the
rotation of the central star. Also, as line three is proportional to
$U^{2}_{\phi}$, its contribution is negligible in the ``suspension
orbit".
   Lastly, two terms of line four in equation (13) have overall positive definite
sign, thus line four serves as the radiation counter-drag. Since
line four is linearly proportional to the cental star's angular
momentum $j$ like the third line, the radiation counter-drag of line
three is also due to the rotation of the central star. However, the
radiation counter-drag of the fourth line acts on the test particle
regardless of the velocity of the particle. Also, they are
comparable to the radiation drag term of line one. Therefore, the
uniform azimuthal velocity of the particle in the ``suspension
orbit" can be attributed to the terms of line four. In other words,
the balance between first line term (radiation drag) and fourth line
terms (radiation counter-drag) makes the particle to hover around
the central star with uniform azimuthal velocity.
  Since the radiation drag term of
the first line exerts on the test particle having the specific
angular momentum $U_{\phi}$ regardless of the spin ($j$) of the
central star,this term exists even when the central star is
not rotating, whereas the radiation counter-drag terms in line two through four
 are linearly proportional to the central star's angular
momentum ($j$), if the central star is non-rotating, all the
radiation counter-drag terms disappear.  Therefore, the emergence of
the radiation counter-drag is attributed to the rotation of the
central star.  It also is of interest that there is another radiation
counter-drag term proportional to the Lense-Thirring angular
velocity $\omega$ and the particle's specific angular momentum
square ($U^{2}_{\phi}$).

   Next, in order to calculate particularly the azimuthal velocity as
measured by a locally static observer, we make use of normalization
condition $g^{\mu\nu}U_{\mu}U_{\nu} = -1$.  That is, plugging $U_{r}
= 0$ of the conditions for the ``suspension orbit" into the
normalization condition yields,
\begin{eqnarray}
       \left(1-\frac{2M}{r}\right) U_{\phi}^{2}
     - 4j\frac{M^2}{r}U_{t}U_{\phi}
     + r^2\left(1-\frac{2M}{r}\right)
     - r^2U_{t}^2 = 0.
\end{eqnarray}

   It is difficult to obtain analytically $U_{\phi}$ and $U_{t}$
from the combination of equations (12) and (14). Therefore, by
assuming that $U_{\phi}^2$ is negligible from our experience in the case
of non-rotating star, i.e., $U_{\phi}^2 \approx 0$ and neglecting
terms of order higher than linear in $j$, we can get approximate
$U_{\phi}$ and $U_{t}$, respectively, as,
\begin{eqnarray}
    U_{\phi} \approx r \left(1-\frac{2M}{r}\right)^{-1/2} \mathcal{J}(r)
                     \left[\frac{B(\alpha)}{A(\alpha)}\right],
\end{eqnarray}
\begin{eqnarray}
    U_{t} \approx -2j\left(\frac{M^2}{r^3}\right) U_{\phi}
                  - \left(1-\frac{2M}{r}\right)^{1/2}.
\end{eqnarray}

   Using $\mathcal{J}(r) \sim j$ in equation (8), equation (15) indicates
$U_{\phi} \sim j$ which is consistent with our assumption above.

   The azimuthal velocity of the test particle as measured by a locally
static observer is given by
\begin{eqnarray}
    v^{\tilde\phi} &=& \frac{U \cdot e^{\tilde\phi}}{U \cdot e^{\tilde t}}
                       \nonumber \\
                   &=& - \frac{1}{r}\left(1-\frac{2M}{r}\right)^{1/2}
                         \left(\frac{U_{\phi}}{U_{t}}\right)
                       + 2j\left(\frac{M^2}{r^2}\right)
                         \left(1-\frac{2M}{r}\right)^{-1/2},
\end{eqnarray}
where $e^{\tilde\phi}$ and $e^{\tilde t}$ are tetrad associated with
the locally static observer. Substitution of $\frac{U_{\phi}}{U_{t}}
\approx
-r\left(1-\frac{2M}{r}\right)^{-1}\mathcal{J}(r)\frac{B(\alpha_{0})}
{A(\alpha_{0})} $ obtained from equations (15) and (16) into
equation (17) yields the azimuthal velocity of the particle in the
``suspension orbit" as measured by the locally static observer to
be,
\begin{eqnarray}
    v_{att}^{\tilde\phi} \approx
                         \left(1-\frac{2M}{r}\right)^{-1/2}
                         \mathcal{J}(r)\frac{B(\alpha_{0})}{A(\alpha_{0})}
                     + 2j\left(\frac{M^2}{r^2}\right)
                         \left(1-\frac{2M}{r}\right)^{-1/2}.
\end{eqnarray}

   First term in equation (18) indicates the contribution to the
azimuthal velocity of the particle in the ``suspension orbit" due to
the combination of the rotation of the star and the rotation of the
radiation source which rotates with the star by being attached to
the stellar surface.  Since second term in equation (18) is equal to
the azimuthal velocity of the LNRF measured by the locally static
observer, it is the contribution due to the Lense-Thirring effect
(dragging of inertial frame) arising from the slow rotation of the
star.

   In order to obtain the coordinate radius from the center of the star to
the ``suspension orbit", we make use of $\frac{U_r}{d\tau} = 0$ and
$U_{r} = 0$ of the conditions characterizing the ``suspension orbit"
together with $U_{\phi}$ in equation (15) and $U_{t}$ in equation
(16).  By inserting these into equation (22) in the Appendix, that
is the $r$-component of the equations of motion, we get
\begin{eqnarray}
    \left(\frac{L^{\infty}}{L^{\infty}_{Edd}}\right) \approx
    \left(1-\frac{2M}{r}\right)^{1/2}
    \left[1-6j\left(\frac{M}{r}\right)\mathcal{J}(r)\frac{B(\alpha)}{A(\alpha)}\right],
\end{eqnarray}
and the solution of equation (19) is the coordinate radius
$r_{suspension}$ from the center of the star to the ``suspension
orbit".
   As we can notice from equation (19), it is noteworthy that the
coordinate radius $r_{so}$ of the ``suspension orbit" indicating the
circular orbit of the particle hovering around the slowly rotating
massive star depends on the luminosity parameter $\left( \frac
{L^{\infty}}{L^{\infty}_{Edd}} \right)$ and the dimensionless
angular momentum $j$ of the star alone, and does not depend on the
initial position and the initial velocity (angular momentum) of the
particles. Also, since equation (21) in the Appendix , that is the
$t$-component of the equations of motion, is satisfied by equations
(12) and (14), the problem of over-determining $U_{t}$ and
$U_{\phi}$ does not occur.
   For $\left( \frac {L^{\infty}}{L^{\infty}_{Edd}}
\right)= 0.75$ and $j=0.10$, equation (19) gives $r_{so} \simeq
4.63M$ which is similar to the value ($r_{so} \simeq 4.66M$) in the
numerical integration of section 3.  Then, plugging $r_{so} \simeq
4.63M$ and $j=0.10$ into equation (18) yields
$v_{suspension}^{\tilde\phi} \simeq 0.056 = 0.044 + 0.012$ which is
also very similar to the value ($v^{\tilde\phi} \simeq 0.056$)
obtained from the numerical integration in section 3, where 0.044 is
the contribution from Doppler shift resulting from the combination
of the rotation of the star and the radiation source, and 0.012 is
the contribution from the Lense-Thirring effect due to the rotation
of the star. Thus, we can find that the Doppler shift due to the
rotation of both the central star and the radiation source have even
larger contribution to the azimuthal velocity of the test particle
than Lense-Thirring effect by the rotation of the central star.

\begin{figure}
\centerline{\epsfig{file=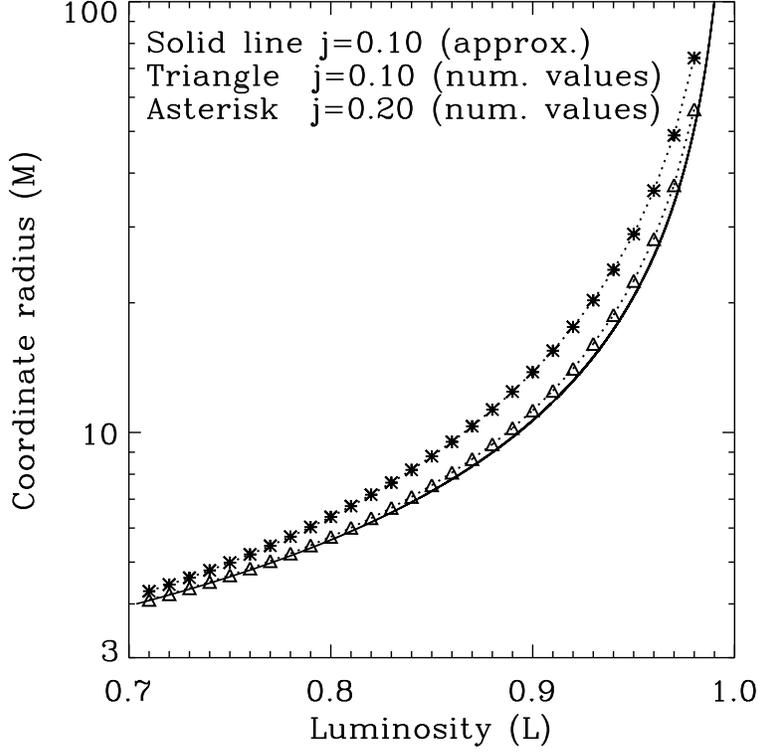, width=10cm, height=10cm}}
\caption{shows the coordinate radius of the ``suspension orbit"
$r_{so}$ as a function of the luminosity $L$ for the angular
momentum of the central star $j = 0.10$ and $0.20$, respectively.
Solid line denotes the plot
 of the approximate expression given by equation (19) for $j=0.10$, and
Triangles ($j=0.10$) and asterisks ($j=0.20$), respectively, denote
values obtained through the numerical integration.},
\end{figure}

   In Fig. 7, as can be seen from the comparison of the solid line (the plot of
the approximate expression given by equation (19)) and numerical
values (triangles and asterisks) obtained by the numerical
integration for $j=0.10$, above approximate equation (19) is valid
only for $j < 0.10$. If $j = 0$ and $v = 0$, i.e., the star is
non-rotating and the radiation source does not rotate either,
equation (19) reduces to,
\begin{eqnarray}
    \left(\frac{L^{\infty}}{L^{\infty}_{Edd}}\right) =
    \left(1-\frac{2M}{r}\right)^{1/2}.
\end{eqnarray}
In this case, for $\left( \frac {L^{\infty}}{L^{\infty}_{Edd}}
\right)= 0.75$, equation (22) gives $r_{so} = 32M/7 \simeq 4.57M$
which is smaller than $r_{so} \simeq 4.66M$ in the slowly rotation
case. Thus, the rotation of the star and the radiation source makes
the ``suspension orbit" to expand outward, which is obvious from the
fact that the rotation of the particles generates centrifugal force
and this force is balanced with the gravitational
force and the radiation pressure.\\

   Also, if $r_{so} = 4M$(stellar radius) and $r_{so} = 6M$ (ISCO)
in the case of $j=0.10$, then from equation (21) we can get
$\left(\frac{L^{\infty}}{L^{\infty}_{Edd}}\right) \simeq 0.70$ and
$\left(\frac{L^{\infty}}{L^{\infty}_{Edd}}\right) \simeq 0.80$,
respectively, so if the luminosity of the star measured at infinity
lies within the range of $0.70 \leq
\left(\frac{L^{\infty}}{L^{\infty}_{Edd}}\right) \leq 0.80$, the
``suspension orbit" appears in boundary layer between the stellar
surface and the ISCO.  Therefore, if the slowly rotating central
star with radius of $R=4M$ has luminosity of $L=0.75$ and angular
momentum of $j=0.10$, the test particle hovers around the star with
azimuthal velocity of about 0.05c in the boundary layer between the
stellar surface and the ISCO where the rotation direction of the
particle is the same as that of the star's spin.

\begin{center}
{\rm\bf V. DISCUSSION}
\end{center}

   We now summarize what we have realized in the present work which
are new ingredients that have not been addressed in the previous
literature.
   The results presented in sections 3 and 4 show that there exists
``suspension orbit" whose location is determined by the central
star's luminosity $L$ as measured at infinity and the star's angular
momentum $j$ and is independent of the initial positions and
velocities of test particles.  To be more concrete, we have explored
the 2-dimensional motion of test particles on the equatorial plane
around slowly rotating star with the asymptotic luminosity measured
at infinity in the range of $0.70 \leq \left (\frac {L^{\infty}}
{L^{\infty}_{Edd}}\right) \leq 1$.  We found out that for the
luminosity in the above range, there exist ``suspension orbit" at
which the radial velocity of the test particle vanishes and proper
time rate of change in the radial and azimuthal velocity also vanish
and the particle's azimuthal velocity $U_{\phi}$ is constant, thus
the test particle hovers around the star with uniform azimuthal
velocity regardless of their initial positions and velocities.
 And it is interesting to note that not only the
radiation drag but also the radiation counter-drags which result
from the central star's spin exert on the particle in the
``suspension orbit" and the balance between the radiation drag and
the radiation counter-drags makes the particle to hover around the
star at uniform azimuthal velocity much greater than that due to the
Lense-Thirring effect (i.e., the dragging of inertial frame).
Furthermore, it is noticeable that there exists another radiation
counter-drag term which is proportional to the Lense-Thirring
angular velocity $\omega$.

   Interestingly enough, Bini, Jantzen, and Stella \cite{DRL09}
reported a study which happens to be akin to the motivation of our
work and consistent with our results in some respects.  They found
the existence of the critical radius at which the test particle
co-rotates with the geometry in the Kerr spacetime background.
However, the radiation stress-energy tensor they employed is valid
for only the photons in outward radial motion with zero angular
momentum, thus is not appropriate to the applications to the
accretion process onto the rotating relativistic stars like the
neutron star.  In the present study, on the other hand, since we
employed the radiation stress-energy tensor first constructed by
Miller and Lamb \cite{ML96} which has no limitations whatsoever on
the character of the emitted photons, our results can be applied to
the accretion process onto the rotating relativistic stars.

   Lastly, in the forthcoming article, we will report on our result of the study
where the central luminous star is non-rotating ($j=0$).

\vspace*{1cm}

\begin{center}
{\rm\bf Acknowledgements}
\end{center}
Authors of this manuscript greatly appreciate the referee's suggestions that will improve the paper very much.
    This work was supported by the Korean Research Foundation Grant
No.2006-341-C00018. JSO acknowledges the support BK21 program to SNU.

\newpage

\begin{center}
{\rm\bf Appendix}
\end{center}

The equations of motion (5) of a tensorial form is decomposed into
each components as follows:

\begin{eqnarray}
   \frac{dU_t}{d\tau} &=& \frac{\sigma}{m}\left(1-\frac{2M}{r}\right)T^{t\beta}U_{\beta}
                       + 2j\frac{\sigma}{m}\left(\frac{M^2}{r}\right)T^{\phi\beta}U_{\beta}
                       - \frac{\sigma}{m}U_{t}T^{\alpha\beta}U_{\alpha}U_{\beta}
                       \\
                      &=& \frac{M}{3R^2}
                          \left( \frac {L^{\infty}}{L^{\infty}_{Edd}} \right)
                          \frac {(1-2M/R)}{(1-2M/r)^2}A(\alpha_{0}) U_{t}
                          \nonumber \\
                      &+& \frac{M}{R^2}
                          \left( \frac {L^{\infty}}{L^{\infty}_{Edd}} \right)
                          \frac {(1-2M/R)}{(1-2M/r)} \sin^2 \alpha_{0} U_{r}
                          \nonumber \\
                      &+& \frac{M}{3R^2}
     R)}{                     \left( \frac {L^{\infty}}{L^{\infty}_{Edd}} \right)
                          \frac {(1-2M/R)}{(1-2M/r)^2}
                          \left[ 2j\left(\frac{M^2}{r^3}\right)
                          A(\alpha_{0}) + \left(\frac{1}{r}\right) \mathcal{J}(r)
                          B(\alpha_{0})\right] U_{\phi} \nonumber \\
                      &-& \frac{M}{3R^2}
                          \left( \frac {L^{\infty}}{L^{\infty}_{Edd}} \right)
                          \frac {(1-2M/(1-2M/r)^3}
                          A(\alpha_{0}) U_{t}^3 \nonumber \\
                      &-& \frac{M}{R^2}
                          \left( \frac {L^{\infty}}{L^{\infty}_{Edd}} \right)
                          \frac {(1-2M/R)}{(1-2M/r)^2}
                          \left[ (2\sin^2\alpha_{0}) U_{t}
                        + \left(1-\frac{2M}{r}\right)(\cos\alpha_{0}\sin^2\alpha_{0}) U_{r}
                          \right] U_{t} U_{r} \nonumber \\
                      &-& \frac{M}{3R^2}
                          \left( \frac {L^{\infty}}{L^{\infty}_{Edd}} \right)
                          \frac {(1-2M/R)}{(1-2M/r)^3}
                          \left[
                          4j\left(\frac{M^2}{r^3}\right) A(\alpha_{0})
                        + \left( \frac{2}{r} \right)\mathcal{J}(r)
                          B(\alpha_{0})\right] U_{t}^2 U_{\phi} \nonumber \\
                      &-& \frac{M}{R^2}
                          \left( \frac {L^{\infty}}{L^{\infty}_{Edd}} \right)
                          \frac {(1-2M/R)}{(1-2M/r)^2}
                          \left[ 4j\left( \frac{M^2}{r^3} \right)
                          (\sin^2\alpha_{0})
                        + \left( \frac{1}{2r} \right)\mathcal{J}(r)
                          (\sin^4\alpha_{0})\right] U_{t} U_{r}
                          U_{\phi},
                          \nonumber
\end{eqnarray}
\begin{eqnarray}
   \frac{dU_r}{d\tau} &=& - \frac{1}{r} +
                            \frac{1}{r}\left(1 - \frac{2M}{r}\right)^{-2}
                            \left(1 - \frac{3M}{r}\right)U^{2}_{t}
                          - \frac{1}{r}\left(1 - \frac{M}{r}\right)U_{r}^2  \\
                      &-&   2j\left(\frac{M^2}{r^4}\right)
                            \left( 1-\frac{2M}{r}\right)^{-2} U_{t} U_{\phi}
                          - \frac{\sigma}{m}\left(1 - \frac{2M}{r}\right)^{-1}
                            T^{r\beta}U_{\beta}
                          - \frac{\sigma}{m}U_{r}T^{\alpha\beta}U_{\alpha}U_{\beta}
                            \nonumber \\
                      &=& - \frac{1}{r} +
                            \frac{1}{r}\left(1 - \frac{2M}{r}\right)^{-2}
                            \left(1 - \frac{3M}{r}\right)U^{2}_{t}
                          - \frac{1}{r}\left(1 - \frac{M}{r}\right)U_{r}^2
                            \nonumber  \\
                      &-&   2j\left(\frac{M^2}{r^4}\right)
                            \left( 1-\frac{2M}{r}\right)^{-2} U_{t} U_{\phi}
                            \nonumber \\
                      &-&   \frac{M}{R^2}
                            \left( \frac{L^{\infty}}{L^{\infty}_{Edd}} \right)
                            \frac{(1-2M/R)}{(1-2M/r)^3} \sin^2\alpha_{0} U_{t}
                            \nonumber \\
                      &-&  \frac{M}{R^2}
                           \left( \frac {L^{\infty}}{L^{\infty}_{Edd}} \right)
                           \frac {(1-2M/R)}{(1-2M/r)^2}
                           (\cos\alpha_{0} \sin^2\alpha_{0}) U_{r}
                           \nonumber \\
                      &-&  \frac{M}{R^2}
                           \left( \frac {L^{\infty}}{L^{\infty}_{Edd}} \right)
                           \frac {(1-2M/R)}{(1-2M/r)^3}
                           \left[ 2j\left(\frac{M^2}{r^3}\right)
                           (\sin^2\alpha_{0})
                         + \left(\frac{1}{4r}\right) \mathcal{J}(r)
                           (\sin^4\alpha_{0})\right] U_{\phi}
                           \nonumber \\
                      &-&  \frac{M}{3R^2}
                           \left( \frac {L^{\infty}}{L^{\infty}_{Edd}} \right)
                           \frac {(1-2M/R)}{(1-2M/r)^3}
                           A(\alpha_{0}) U_{t}^2 U_{r}
                           \nonumber \\
                      &-&  \frac{M}{R^2}
                           \left( \frac {L^{\infty}}{L^{\infty}_{Edd}} \right)
                           \frac {(1-2M/R)}{(1-2M/r)^2}
                           \left[(2\sin^2\alpha_{0}) U_{t}
                         + \left(1-\frac{2M}{r}\right)(\cos\alpha_{0}\sin^2\alpha_{0})U_{r}
                           \right]U_{r}^2
                           \nonumber \\
                      &-&  \frac{M}{3R^2}
                           \left( \frac {L^{\infty}}{L^{\infty}_{Edd}} \right)
                           \frac {(1-2M/R)}{(1-2M/r)^3}
                           \left[ 4j\left(\frac{M^2}{r^3}\right) A(\alpha_{0})
                         + \left( \frac{2}{r} \right)\mathcal{J}(r)B(\alpha_{0})
                           \right] U_{t} U_{r} U_{\phi}
                           \nonumber \\
                      &-&  \frac{M}{R^2}
                           \left( \frac {L^{\infty}}{L^{\infty}_{Edd}} \right)
                           \frac {(1-2M/R)}{(1-2M/r)^2}
                           \left[ 4j\left( \frac{M^2}{r^3} \right)(\sin^2\alpha_{0})
                         + \left( \frac{1}{2r} \right)\mathcal{J}(r)(\sin^4\alpha_{0})
                           \right] U_{r}^2 U_{\phi},
                           \nonumber
\end{eqnarray}
\begin{eqnarray}
   \frac{dU_{\phi}}{d\tau} &=&
      2j\frac{\sigma}{m}\left(\frac{M^2}{r}\right)T^{t\beta}U_{\beta}
    - \frac{\sigma}{m}r^2 T^{\phi\beta}U_{\beta}
    - \frac{\sigma}{m}U_{\phi}T^{\alpha\beta}U_{\alpha}U_{\beta} \\
                           &=& - \frac{M}{3R^2}
                          \left( \frac {L^{\infty}}{L^{\infty}_{Edd}} \right)
                          \frac {(1-2M/R)}{(1-2M/r)^3} r \mathcal{J}(r)B(\alpha_{0}) U_{t}
                          \nonumber \\
                      &-& \frac{M}{R^2}
                          \left( \frac {L^{\infty}}{L^{\infty}_{Edd}} \right)
                          \frac {(1-2M/R)}{(1-2M/r)^2}
                          \left(\frac{r}{4}\right) \mathcal{J}(r)
                          (\sin^4\alpha_{0}) U_{r}
                          \nonumber \\
                      &-& \frac{M}{3R^2}
                          \left( \frac {L^{\infty}}{L^{\infty}_{Edd}} \right)
                          \frac {(1-2M/R)}{(1-2M/r)^3}A(\alpha_{0}) U_{t}^2 U_{\phi}
                          \nonumber \\
                      &-& \frac{M}{R^2}
                          \left( \frac {L^{\infty}}{L^{\infty}_{Edd}} \right)
                          \frac {(1-2M/R)}{(1-2M/r)^2}
                          \left[ (2\sin^2\alpha_{0})U_{t}
                        + \left(1-\frac{2M}{r}\right) (\cos\alpha_{0}\sin^2\alpha_{0})U_{r}
                          \right] U_{r}U_{\phi}
                          \nonumber \\
                      &-& \frac{M}{3R^2}
                          \left( \frac {L^{\infty}}{L^{\infty}_{Edd}} \right)
                          \frac {(1-2M/R)}{(1-2M/r)^3}
                          \left[ 4j\left( \frac{M^2}{r^3} \right) A(\alpha_{0})
                        + \left( \frac{2}{r} \right)\mathcal{J}(r)B(\alpha_{0})
                          \right] U_{t} U_{\phi}^2 \nonumber \\
                      &-& \frac{M}{R^2}
                          \left( \frac {L^{\infty}}{L^{\infty}_{Edd}} \right)
                          \frac {(1-2M/R)}{(1-2M/r)^2}
                          \left[ 4j\left( \frac{M^2}{r^3} \right)(\sin^2\alpha_{0} )
                        + \left( \frac{1}{2r} \right)\mathcal{J}(r)(\sin^4\alpha_{0})
                          \right] U_{r} U_{\phi}^2, \nonumber
\end{eqnarray}
where $A(\alpha_{0}) = \cos^3\alpha_{0} -9\cos\alpha_{0} + 8$,
$B(\alpha_{0}) = \cos^3\alpha_{0} -3\cos\alpha_{0} + 2$, and
$\sin\alpha_{0} = \left(\frac{R}{r} \right) {\left( \frac{1 -
2M/r}{1 - 2M/R}\right)}^{1/2}$.

\end{document}